\documentclass[conference,9pt]{IEEEtran}
\IEEEoverridecommandlockouts

\usepackage{cite}
\usepackage{amsmath,amssymb,amsfonts}
\usepackage{algorithmic}
\usepackage{graphicx}
\usepackage{textcomp}
\usepackage{xcolor}

\usepackage{booktabs}
\usepackage[hidelinks]{hyperref}
\usepackage{siunitx}
\usepackage{mathtools}
\usepackage{comment}
\usepackage{multirow}
\usepackage{makecell}

\makeatletter
\let\MYcaption\@makecaption
\makeatother
\usepackage[font=footnotesize,subrefformat=parens]{subcaption}
\makeatletter
\let\@makecaption\MYcaption
\makeatother

\usepackage{tikz}
\usetikzlibrary{calc}
\usetikzlibrary{tikzmark}
\usetikzlibrary{arrows, positioning,intersections,angles}
\usetikzlibrary{fit}
\usetikzlibrary{patterns}

\usepackage{xspace}
\makeatletter
\DeclareRobustCommand\onedot{\futurelet\@let@token\@onedot}
\def\@onedot{\ifx\@let@token.\else.\null\fi\xspace}

\makeatother

\def\equationautorefname~#1\null{(#1\null)}

\renewcommand{\sectionautorefname}{Section}
\renewcommand{\subsectionautorefname}{\sectionautorefname}

\let\orgautoref\autoref

\renewcommand{\autoref}[1]
{%
\def\figureautorefname{Fig.}%
\def\subfigureautorefname{\figureautorefname}%
\def\sectionautorefname{Sec.}%
\def\subsectionautorefname{\sectionautorefname}%
\def\subsectionautorefname{\sectionautorefname}%
\orgautoref{#1}%
}

\newcommand{\vect}[1]{\mbox{\boldmath $#1$}}
\newcommand{\abs}[1]{\left\lvert#1\right\rvert}

\newcommand{\trans}[1]{#1^\mathsf{T}}

\def\appendixautorefname~#1\null{~#1 \null}

\newcommand{\svss}{\textit{one-vs-one}}
\newcommand{\svsm}{\textit{one-vs-many}}

\makeatletter
\newcommand{\figcaption}[1]{\def\@captype{figure}\caption{#1}}
\newcommand{\tblcaption}[1]{\def\@captype{table}\caption{#1}}
\makeatother

\makeatletter 
\newcommand{\linebreakand}{%
  \end{@IEEEauthorhalign}
  \hfill\mbox{}\par
  \mbox{}\hfill\begin{@IEEEauthorhalign}
}
\makeatother 

\def\BibTeX{{\rm B\kern-.05em{\sc i\kern-.025em b}\kern-.08em
    T\kern-.1667em\lower.7ex\hbox{E}\kern-.125emX}}
    
\begin{document}
\abovedisplayskip=1.3ex plus 3pt minus 1pt
\belowdisplayskip=\abovedisplayskip

\setlength\abovecaptionskip{4pt}
\setlength\abovetopsep{-7pt}

\title{Guided Speaker Embedding}

\author{\IEEEauthorblockN{Shota Horiguchi, Takafumi Moriya, Atsushi Ando, Takanori Ashihara, Hiroshi Sato, Naohiro Tawara, and Marc Delcroix}
\IEEEauthorblockA{NTT Corporation, Japan}}

\maketitle

\begin{abstract}
This paper proposes a guided speaker embedding extraction system, which extracts speaker embeddings of the target speaker using speech activities of target and interference speakers as clues.
Several methods for long-form overlapped multi-speaker audio processing are typically two-staged: i) segment-level processing and ii) inter-segment speaker matching.
Speaker embeddings are often used for the latter purpose.
Typical speaker embedding extraction approaches only use single-speaker intervals to avoid corrupting the embeddings with speech from interference speakers.
However, this often makes speaker embeddings impossible to extract because sufficiently long non-overlapping intervals are not always available.
In this paper, we propose using speaker activities as clues to extract the embedding of the speaker-of-interest directly from overlapping speech.
Specifically, we concatenate the activity of target and non-target speakers to acoustic features before being fed to the model.
We also condition the attention weights used for pooling so that the attention weights of the intervals in which the target speaker is inactive are zero.
The effectiveness of the proposed method is demonstrated in speaker verification and speaker diarization.
\end{abstract}

\begin{IEEEkeywords}
speaker embedding, speaker verification, speaker diarization
\end{IEEEkeywords}

\section{Introduction}
Dealing with multi-speaker overlapping long-form speech is one of the ultimate challenges in speech processing.
One large trend in long-form multi-speaker speech processing is to first process each segmented short signal of possible including overlapped speech, and then integrate the segment-wise results.
For example, in multi-speaker automatic speech recognition (ASR), some transducer-based methods are designed to transcribe each speech segment extracted from a long-form recording~\cite{lu2021streaming,lu2022endpoint,sklyar2022multi,kanda2022streaming,moriya2025alignment}.
These methods offer rough time stamps of each recognized token with intra-segment speaker labels.
To obtain a result for the long-form recording, inter-segment speaker labels are estimated by referring to similarities of speaker embeddings extracted for each speaker in each segment.
Also in speaker diarization, EEND-vector clustering (EEND-VC)~\cite{kinoshita2021integrating,kinoshita2021advances} and pyannote~\cite{bredin2023pyannote,plaquet2023powerset} first perform segment-wise overlap-aware speaker diarization, and then obtain inter-segment results on the basis of similarities between speaker embeddings.
Some methods support joint training of segment-level speech processing and speaker embedding extraction~\cite{kinoshita2021integrating,kinoshita2021advances,lu2021streaming2,kanda2022streaming2}, but using an external speaker embedding extractor has been reported to improve performance~\cite{tawara2024ntt}.
This is because speaker embedding extractors can benefit from large-scale noisy datasets covering many speakers like VoxCeleb~\cite{nagrani2020voxceleb} for training, while multi-speaker speech tasks require well-annotated clean datasets for simulating mixtures, which limits the variation of speakers.
Therefore, speaker embedding extraction from (partially) overlapped speech is the key to these applications.

In general, to obtain reliable speaker embeddings, the input speech should preferably be as long as possible~\cite{snyder2017deep}.
However, since speaker embedding extractors are generally made to extract speaker embeddings from single-speaker recordings~\cite{snyder2017deep,snyder2018xvectors,desplanques2020ecapatdnn,zhou2021resnext}, there is a trade-off between duration and purity in the case of overlapped speech.
Using overlapped intervals allows for longer speech but lowers the purity of the speaker, resulting in poor-quality speaker embeddings.
In contrast, using only single-speaker intervals results in short input speech, and in the worst case, it is not available.
To obtain reliable speaker embeddings, it is necessary to fully utilize the available overlapping speech segments, not limited to single-speaker intervals.

Several studies have shown that speaker embeddings can be extracted even from fully overlapped speech~\cite{han2020mirnet,cord2023teacher,horiguchi2024recursive}; these methods can be promising solutions for extracting an embedding for each speaker that appeared in a segment in the aforementioned multi-speaker applications.
Especially, a very recent study has shown that this can be achieved by computing speaker-wise attention weights used in the pooling step~\cite{horiguchi2024recursive}.
This means that if we can compute attention weights properly for each speaker in a segment, it is highly possible to selectively extract an embedding for each of them without significantly modifying network architecture.
However, since the conventional method is just designed to extract embeddings for all the speakers appearing in the input speech, it is not obvious which one corresponds to each speaker in a segment.
To use speaker embeddings with these multi-speaker applications, each embedding must be associated with intra-segment speaker labels.
In other words, in each segment, we need to first specify the speaker of interest, i.e., the target speaker, and then extract an embedding for the speaker.

\begin{figure}[t]
\centering
\input{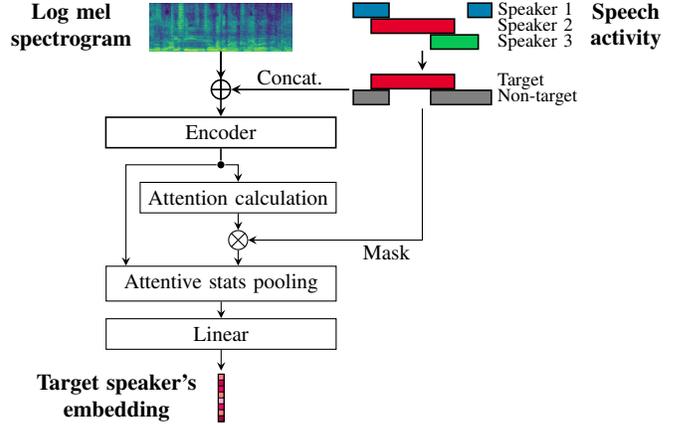}
\caption{Schematic diagram of guided speaker embedding (Target: Speaker 2).}\label{fig:diagram}
\end{figure}

In this paper, we propose a \textit{guided speaker embedding} extraction, a simple but effective method to extract the target speaker's embedding.
As shown in \autoref{fig:diagram}, in the proposed method, the target speaker is conditioned by using speech activities, which are naturally obtained in the aforementioned multi-speaker applications as segment-wise results.
The activity of the target speaker is used to avoid attending to the intervals in which the target speaker is inactive.
Furthermore, feeding the activities of target and non-target speakers into the model in addition to the log mel spectrogram enables the model to suppress the information of interference speakers with the encoder.
This structure gives the proposed method three advantages: i) it enables an embedding of the target speaker to be extracted, ii) it does not rely on the number of interference speakers, and iii) it can follow the standard training protocol of speaker embedding extractors without introducing auxiliary tasks such as ASR or diarization.
The experimental evaluation was conducted on two tasks: speaker verification and speaker diarization.
Speaker verification is to show the discriminativeness of the embeddings extracted using the proposed method, and speaker diarization is to show the effectiveness of the proposed method in a practical use case.

\section{Related work}
Speaker activity information plays an essential role in extracting speaker embeddings from multi-speaker recordings.
The most naive approach is to extract speaker embeddings from the segments in which each speaker is speaking alone.
This approach is used in many two-staged speaker diarization methods \cite{kinoshita2021integrating,kinoshita2021advances,bredin2023pyannote,plaquet2023powerset,taherian2024multi}.
There are also methods that iteratively perform target-speaker speech processing to obtain speaker-wise speech activity and speaker embedding extraction from the estimated single-speaker intervals \cite{kanda2019simultaneous,medennikov2020targetspeaker}.
However, avoiding overlapped intervals may lead to poor performance when the single-speaker intervals are not long enough.

Besides, more effective ways to use speech activity have also been proposed.
One successful approach is guided source separation (GSS), which uses speaker diarization results as prior information to estimate speaker-wise separation mask~\cite{boeddeker2018front}.
This technique has become a \textit{de facto} standard in CHiME-6~\cite{watanabe2020chime}, which is a typical example of wild data, indicating the potential of utilizing speech activity for processing overlapping speech.
As a neural-network-based method, a memory-aware multi-speaker embedding mechanism extracts embeddings of all the speakers included in diarization results utilizing overlapped intervals~\cite{he2023ansdmamse,yang2024neural}.
Since it is a submodule of the target-speaker voice activity detector, the resulting embeddings are not designed to be speaker discriminative, i.e., the model cannot be used as an off-the-shelf speaker embedding extractor.
Another example is a speaker activity driven speech extraction neural network (ADEnet), which uses the speaker activity of the target speaker for speaker extraction~\cite{delcroix2021speaker}.
It has the potential to improve performance by using information from the interference speakers as in GSS.
Our proposed method provides an off-the-shelf speaker embedding extractor by utilizing information from both target and interference speakers.

\section{Method}
\subsection{Conventional method: Basic speaker embedding}
We first briefly introduce the basic processing flow in a speaker embedding extractor.
Given an $L$-length sequence of $F$-dimensional acoustic features $\left[\vect{x}_1,\dots,\vect{x}_L\right]\in\mathbb{R}^{F\times L}$, it is first input to the encoder $f$ to obtain $T$-length $D$-dimensional frame-wise embeddings $\left[\vect{h}_1,\dots,\vect{h}_T\right]\in\mathbb{R}^{D\times T}$:
\begin{equation}
    \left[\vect{h}_1,\dots,\vect{h}_T\right]=f\left(\vect{x}_1,\dots,\vect{x}_{L}\right).
\end{equation}
Whether the sequence length changes depends on the encoder, and for ECAPA-TDNN~\cite{desplanques2020ecapatdnn} used in this paper, $L=T$.
The frame-wise embeddings from the encoder are then aggregated into a single embedding using a pooling module.
In this paper, we use channel- and context-wise attentive statistics pooling~\cite{desplanques2020ecapatdnn}, in which attention weights $\left[\vect{a}_1,\dots,\vect{a}_T\right]\in\left(0,1\right)^{D\times T}$ used for pooling are computed as follows:
\begin{align}
    \left[\vect{a}_1,\dots,\vect{a}_T\right]&=\mathsf{Softmax}\left(\left[\tilde{\vect{a}}_1,\dots,\tilde{\vect{a}}_T\right]\right),\label{eq:attn}\\
    \tilde{\vect{a}}_t&=W_2 g\left(W_1\vect{e}_t+\vect{b}_1\right)+\vect{b}_2\in\mathbb{R}^D.\label{eq:attn_logit}
\end{align}
Here, $\vect{e}_t\coloneqq\vect{h}_t\oplus\vect{\mu}\oplus\vect{\sigma}\in\mathbb{R}^{3D}$, where $\vect{\mu}$ and $\vect{\sigma}$ are the element-wise mean and standard deviation of $\left[\vect{h}_1,\dots,\vect{h}_T\right]$, $\oplus$ is vector concatenation, $W_1$ and $\vect{b}_1$ ($W_2$ and $\vect{b}_2$) are the weight and bias of the first (second) linear layer, $g\left(\cdot\right)$ is the rectified linear unit, and $\mathsf{Softmax}\left(\cdot\right)$ is the row-wise softmax normalization.
With the computed attention weights, the speaker embedding $\vect{v}\in\mathbb{R}^{E}$ is obtained via
\begin{align}
    \vect{v}=W_o\left(\tilde{\vect{\mu}}\oplus\tilde{\vect{\sigma}}\right)+\vect{b}_o,\label{eq:embed}
\end{align}
where $W_o\in\mathbb{R}^{E\times 2D}$ and $\vect{b}_o\in\mathbb{R}^E$ are the weight and bias of the last linear layer, and $\tilde{\vect{\mu}}$ and $\tilde{\vect{\sigma}}$ are the element-wise weighted mean and standard deviation:
\begin{align}
    \tilde{\vect{\mu}}&=\sum_{\tau=1}^{T}\vect{a}_{\tau}\odot\vect{h}_{\tau}\in\mathbb{R}^D,\label{eq:embed_mean}\\
    \tilde{\vect{\sigma}}&=\sqrt{\sum_{\tau=1}^{T}\vect{a}_{\tau}\odot\vect{h}_{\tau}\odot\vect{h}_{\tau}-\tilde{\vect{\mu}}\odot\tilde{\vect{\mu}}}\in\mathbb{R}^D,\label{eq:embed_std}
\end{align}
where $\odot$ denotes the Hadamard product.

\subsection{Proposed method: Guided speaker embedding}
Since the conventional method introduced in the previous section aggregates the frame-wise embeddings into a single embedding, the input is assumed to contain only a single speaker.
Some conventional methods tackled extracting speaker-wise embeddings from multi-speaker recordings \cite{han2020mirnet,cord2023teacher,horiguchi2024recursive}, but they do not support extracting embeddings of the speaker-of-interest.

While ADEnet only uses the target speaker's speech activity~\cite{delcroix2021speaker}, the presence of interference speakers does not ensure that all the target speaker's speech intervals are useful.
As in GSS~\cite{boeddeker2018front}, intervals in which interference speakers speak are also helpful in finding the target speaker's information since they include which information the target speaker's embedding should not contain.
One possible way is to use all speakers' speech activities directly by concatenating them to the input feature.
However, the number of speakers assumed in a segment differs among applications, e.g., pyannote assumes at most three speakers per \SI{10}{\second}~\cite{bredin2023pyannote}, EEND-VC assumes three speakers per \SI{30}{\second}~\cite{kinoshita2021advances}, and token-level serialized output training assumes two or five speakers in a segment~\cite{kanda2022streaming}; thus, fixing the network architecture to a specific number of speakers would significantly reduce its practicality.

To handle arbitrary numbers of interference speakers, we instead use the joint activity of interference speakers, i.e., non-target speakers' activity, as shown in the top right of \autoref{fig:diagram}.
Let $\vect{z}_l\coloneqq\trans{\left[z_{1,l},\dots,z_{N,l}\right]}\in\left\{0,1\right\}^{N}$ be $N$ speakers' speech activities at $l$, where $z_{n,l}=1$ if the $n$-th speaker is active at $l$ and $0$ otherwise, and $m\in\left\{1,\dots,N\right\}$ be the target speaker's index.
In the proposed method, the input to the encoder $\left[\vect{x}'_1,\dots,\vect{x}'_L\right]$ is extended from the original acoustic feature $\left[\vect{x}_1,\dots,\vect{x}_L\right]$ as 
\begin{equation}
    \vect{x}'_l=\vect{x}_l\oplus\trans{\left[z_l^{\text{target}},z_l^{\text{non-target}}\right]}\in\mathbb{R}^{F+2},
\end{equation}
where $z_l^{\text{target}}\coloneqq z_{m,l}$ and $z_l^{\text{non-target}}$ is defined as follows:
\begin{equation}
    z_l^{\text{non-target}}\coloneqq\begin{cases}
        1 & \left(\text{if}~\exists n\in\left\{1,\dots,N\right\}\setminus\left\{m\right\}, z_{n,l} = 1\right),\\
        0 & (\text{otherwise}).
    \end{cases}
\end{equation}
This enables the method to work with an arbitrary number of interference speakers.

We also know that the information at frames in which the target speaker is inactive should not be carried to the speaker embedding calculation.
For the calculation of $\vect{\mu}$ and $\vect{\sigma}$ to be used for $\vect{e}_t$, only frames in which the target speaker is active are used.
We also zero out the attention weight at the frames by using the target speaker's activity as a mask and re-normalize the attention weights in the intervals where the target speaker is active:
\begin{equation}
    \vect{a}_t\leftarrow
    \begin{cases}
        \vect{a}_t\oslash\sum_{\tau=1}^T z_{\left\lfloor{L\tau/T}\right\rceil}^{\text{target}}\vect{a}_{\tau} & (\text{if}~z_{\left\lfloor{L\tau/T}\right\rceil}^{\text{target}}=1),\\
        \smash{\trans{[\underbrace{0,\dots,0}_{D}]}} & (\text{otherwise}).
    \end{cases}
    \vphantom{\begin{cases}
        \vect{a}_t\oslash\sum_{\tau=1}^T\vect{a}_{\tau} & (\text{if}~z_{n,t}=1),\\
        \trans{[\underbrace{0,\dots,0}_{N}]} & (\text{otherwise}),
    \end{cases}}
    \label{eq:attention_masking}
\end{equation}
where $\oslash$ denotes the Hadamard division and $\lfloor\cdot\rceil$ is a rounding function.
The updated attention weights are used for attentive statistics pooling in \autoref{eq:embed_mean}--\autoref{eq:embed_std}.

\section{Experimental setup}
Our speaker embedding extractor was based on ECAPA-TDNN with 1\,024 channels~\cite{desplanques2020ecapatdnn}.
The model takes 80-dimensional log mel filterbank features extracted using a sliding window of \SI{25}{\ms} width and \SI{10}{\ms} shift as input.
In the case of the proposed method, target and non-target speech activities were also fed to the model, resulting in 82-dimensional input.
The dimensionality of the output speaker embeddings was set to 192.
The baseline ECAPA-TDNN was trained using single-speaker utterances cropped to \SI{3}{\second} with a mini-batch size of 256.
The proposed model was trained using 128 three-speaker partially-overlapped mixtures generated on the fly at each iteration, resulting in 384 samples in a minibatch.
Each mixture was generated using utterances cropped to 3--6 seconds with the constraint that the start times of each utterance differ by at least \SI{0.5}{\second}, which is similar to the way used in a multi-speaker ASR study~\cite{kanda2022streaming}.
The mixture ratio was randomly sampled from $[-5,5]$~\si{\dB}.

The models were trained as multi-class classifiers using additive angular margin softmax loss with margin and scaling parameters of 0.2 and 30, respectively.
The Adam optimizer with cyclical learning rate scheduling for four cycles was used for optimization.
Each cycle consists of 20 epochs, with the first 1k iterations for warm-up and the rest for cosine annealing decay.
The peak learning rate of the first cycle was 0.001 and decayed by 0.75 at each cycle.

As the training dataset, we used the development sets of VoxCeleb 1 and 2, including 1\,240\,651 utterances from 7\,205 speakers.
On-the-fly data augmentation using noise~\cite{snyder2015musan} and reverberation~\cite{ko2017study} are randomly applied during training with each probability of 0.5, respectively.

The evaluation was conducted in both speaker verification and speaker diarization to show that the proposed method is suitable for general-purpose use.
Evaluation details are described in each of the following sections.

\section{Results}
\subsection{Speaker verification}
\begin{figure}[t]
    \centering
    \subfloat[Overlap ratio of the target speaker's utterance]{\includegraphics[width=0.49\linewidth]{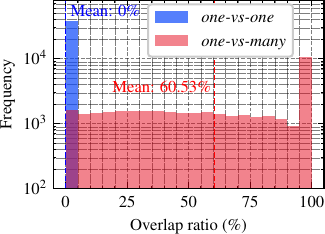}\label{fig:overlap_ratio}}
    \hfill
    \subfloat[Duration of single-speaker intervals]{\includegraphics[width=0.49\linewidth]{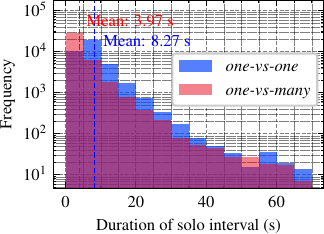}\label{fig:solo_len}}
    \caption{Statistics of target speaker's utterances ($n=37\,611$).}\label{fig:stats}
\end{figure}
Speaker verification performance was evaluated under the standard \svss{} protocol and novel \svsm{} protocol, respectively.
For the \svss{} protocol, VoxCeleb1-O~\cite{nagrani2020voxceleb} was used for evaluation, which consists of 37\,611 trials.
For the \svsm{} protocol, we extended each trial of VoxCeleb1-O by mixing three interference speakers' speech to the evaluation speech with random delays to simulate segment-wise results in multi-speaker applications.
The $n$-th speaker's utterance was arranged to start after $\delta_n$ of the $(n-1)$-th speaker's utterance $u_{n-1}$ starts, where $\delta_n$ is randomly sampled from $\left[0,\abs{{u}_{n-1}}\right]$.
The statistics of the evaluation datasets are shown in \autoref{fig:stats}.
\autoref{fig:overlap_ratio} shows the distribution of overlap ratios of the target speakers.
The evaluation dataset of \svsm{} condition shows a high overlap ratio of \SI{60.53}{\percent}, which is higher than that of typical real conversations.
Note that the target speaker's speech is fully overlapped in 9\,843 mixtures, and the overlap ratios of the partially overlapped mixtures are almost uniformly distributed.
\autoref{fig:solo_len} shows the distribution of durations in which only target speakers' are speaking individually.
Due to the high overlap ratio, the average duration of single-speaker intervals is \SI{3.97}{\second}, which is less than half of that of VoxCeleb1-O.
This is a severe scenario for conventional methods that use only the target speaker's single-speaker intervals for speaker embedding extraction, and extracting informative information from the overlapped intervals is the key to improving the performance.\footnote{When the target speaker's speech was fully overlapped, the overlapped interval was used for extracting speaker embedding, similar to the way used in pyannote \cite{bredin2023pyannote}.}
In the evaluation, the ground-truth activity of each speaker was assumed to be available.
Each method was evaluated using equal error rates (EERs) and minimum detection cost function (minDCF) with $p_\text{target}=0.01$ based on cosine similarity between embeddings.

\begin{table}[t]
    \caption{Speaker verification results in EERs (\%) and minDCF.}\label{tbl:result_veri}
    \centering
    \sisetup{detect-weight,mode=text}
    \setlength{\tabcolsep}{2pt}
    \resizebox{\linewidth}{!}{%
    \begin{tabular}{@{}llS[table-format=1.2]S[table-format=1.2]S[table-format=2.2]S[table-format=1.2]@{}}
        \toprule
        &&\multicolumn{2}{c}{\svss}&\multicolumn{2}{c@{}}{\svsm}\\\cmidrule(l{\tabcolsep}r{\tabcolsep}){3-4}\cmidrule(l{\tabcolsep}){5-6}
        ID&Method&{EER}&{minDCF}&{EER}&{minDCF}\\\midrule
        \multicolumn{5}{@{}l}{\textbf{Single-speaker model (baseline)}}\\
        \texttt{B1}&Input: Single-speaker intervals& 0.88 & 0.09 & 12.84 & 0.36 \\
        \texttt{B2}&Input: Single-speaker \& overlapped intervals& 0.88 & 0.09 & 14.11 & 0.39 \\\midrule
        \multicolumn{5}{@{}l}{\textbf{Guided speaker embedding (proposed)}}\\
        \texttt{P1}& Complete proposed method& 1.09 & 0.11 & 4.04 & 0.22 \\
        \texttt{P2}& $\hookrightarrow$ w/o target-speaker input $z_l^\text{target}$ & 1.29 & 0.14 & 5.46 & 0.28\\
        \texttt{P3}& $\hookrightarrow$ w/o non-target-speaker input $z_l^\text{non-target}$ & 1.24 & 0.13 & 4.53 & 0.25\\
        \texttt{P4}& $\hookrightarrow$ w/o attention masking in \autoref{eq:attention_masking} & 1.71 & 0.19 & 4.83 & 0.30\\
        \bottomrule
    \end{tabular}%
    }
\end{table}

\begin{table*}[t]
    \caption{Speaker diarization results in DERs (\%) / JERs (\%).}
    \label{tbl:result_diar}
    \centering
    \sisetup{detect-weight,mode=text}
    \renewrobustcmd{\bfseries}{\fontseries{b}\selectfont}
    \renewrobustcmd{\boldmath}{}
    \newrobustcmd{\B}{\bfseries}
    \begin{tabular}{@{}lS[table-format=2.2]S[table-format=2.2]@{\ /\ }S[table-format=2.2]*{4}{S[table-format=2.2]@{\ /\ }S[table-format=2.2]}@{}}
    \toprule
    &\multicolumn{1}{c}{\multirow{2.92}{*}{\makecell{Overlap\\ratio (\%)}}}&\multicolumn{2}{c}{\multirow{2.92}{*}{\makecell{Original\\pyannote \cite{plaquet2023powerset}}}}&\multicolumn{4}{c}{Local diarization: pyannote \cite{plaquet2023powerset}}&\multicolumn{4}{c@{}}{Local diarization: Oracle}\\\cmidrule(lr){5-8}\cmidrule(l){9-12}
    Dataset &  & \multicolumn{2}{c}{}&\multicolumn{2}{c}{Baseline (\texttt{B1})} & \multicolumn{2}{c}{Guided (\texttt{P1})} & \multicolumn{2}{c}{Baseline (\texttt{B1})} & \multicolumn{2}{c@{}}{Guided (\texttt{P1})}\\\midrule
    AISHELL-4 \cite{fu2021aishell} & 4.95 &12.22&18.28& 12.44& 19.91 & \B 11.69 & \B 18.86 & \B 3.26 & \B 9.32 & 3.60 & 9.70\\
    AliMeeting \cite{yu2022m2met} & 20.36 &24.38&29.44& 21.52 & 26.38 & \B 20.47 & \B 24.50 & 9.09 & 11.19 & \B 4.57 & \B 7.14\\
    AMI Mix-Headset \cite{carletta2007unleashing} & 14.58 &18.80&23.57& 17.98 & 24.20 & \B 16.80 & \B 22.85 & 5.97 & 10.49 & \B 3.09 & \B 7.52\\
    AMI Array channel 1 \cite{carletta2007unleashing} & 14.58 &22.45&27.84& 23.27 & 30.92 & \B 20.44 & \B 27.09 & 8.30 & 14.40 & \B 7.25 & \B 10.50\\
    MSDWild (few) \cite{liu2022msdwild} & 12.43 &24.93&47.16& 25.85 & 46.53 & \B 25.27 & \B 44.97 & \B 10.52 & 26.27 & 10.54 & \B 25.26\\
    MSDWild (many) \cite{liu2022msdwild} & 17.17 &46.03&75.12& 50.10 & 76.94 & \B 49.64 & \B 75.24 & 30.42 & 60.07 & \B 29.75 & \B 58.50\\
    DIHARD III \cite{ryant2021third} & 9.37 & 21.66 & 38.22 & \B 21.07 & 39.28 & 22.33 &\B 38.94 & \B 6.27 & 24.67 & 6.46 & \B 24.28\\
    VoxConverse \cite{chung2020spot} & 3.05 &11.29&34.23& \B 11.48 & 35.80 & 12.09 & \B 32.91 & 2.77 & 23.36 & \B 2.40 & \B 22.48\\\midrule
    Macro average &&22.72&36.73&22.96&37.50&\B 22.34&\B 35.67&9.58& 22.47& \B 8.46& \B 20.67\\
    \bottomrule
    \end{tabular}
\end{table*}

The results are shown in \autoref{tbl:result_veri}.
In terms of the conventional method, using single-speaker intervals showed \SI{0.88}{\percent} EER and $0.09$ minDCF in the \svss{} scenario, but severely degraded performance in the \svsm{} scenario (\texttt{B1}).
Using overlaps in addition showed the same good performance in the \svss{} scenario because there were no overlaps, and the performance of \svsm{} was more degraded (\texttt{B2}).
The proposed guided speaker embedding drastically improved the performance of the \svsm{} scenario with a small degradation in \svss{} (\texttt{P1}); this clearly indicates the effectiveness of the proposed method.
The last three rows show the results of the ablation study by removing input feature/component from \texttt{P1} in both the training and inference stages.
Lacking the target speaker's activity in input severely degraded the performance of \svsm{} (\texttt{P2}); this is because the model was trained to avoid using intervals in which non-target speakers are active, resulting in poor performance on highly or fully overlapped cases.
The method lacking non-target speakers' activity in input still performed reasonably well (\texttt{P3}), but was not as well as the proposed method.
The reason is that the intervals of the target speaker are unknown, and thus, for example, interference speakers that fully overlap the target speaker cannot be detected.
The verification performance was degraded when attention masking was not used, especially in the \svss{} scenario (\texttt{P4}).
This might be because the attention needed to be more aggressive to filter out non-target speakers' intervals, resulting in over-deletion of the target speaker's information.

\begin{figure}
    \centering
    \subfloat[Partial overlap]{%
        \begin{minipage}{0.48\linewidth}
        \includegraphics[width=\linewidth]{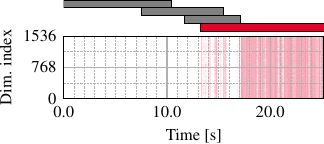}\vspace{0.5em}
        \includegraphics[width=\linewidth]{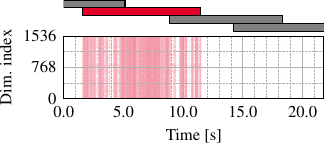}
        \end{minipage}%
    }\hfill
    \subfloat[Full overlap]{%
        \begin{minipage}{0.48\linewidth}
        \includegraphics[width=\linewidth]{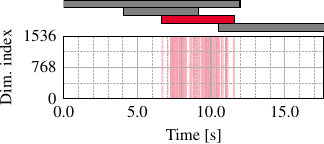}\vspace{0.5em}
        \includegraphics[width=\linewidth]{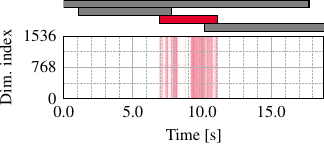}
        \end{minipage}%
    }
    \caption{Examples of attention weights calculated using the proposed method. Red and gray bars visualize the speech activities of target and interference speakers, respectively.}\label{fig:vis_attn}
\end{figure}

\autoref{fig:vis_attn} visualized the attention weights of the partially and fully overlapped cases.
In the partially overlapped case, single-speaker intervals generally have larger attention weights, but overlap intervals clearly also attract attention.
In the fully overlapped case, selective attention is applied.

\subsection{Speaker diarization}
Speaker diarization performance was evaluated using the following datasets: AISHELL-4~\cite{fu2021aishell}, AliMeeting~\cite{yu2022m2met}, AMI Mix-Headset and the first channel of the first array~\cite{carletta2007unleashing}, MSDWild~\cite{liu2022msdwild}, DIHARD III~\cite{ryant2021third}, and VoxConverse~\cite{chung2020spot}. 
The evaluation was conducted on the basis of pretrained pyannote 3.1~\cite{bredin2020pyannote,plaquet2023powerset}, in which overlap-aware speaker diarization is performed for a \SI{10}{\second} sliding window with \SI{1}{\second} shift.
Speaker embeddings extracted for active speakers in each window are used to find the global speaker labels on the basis of their similarity.
The original pyannote uses single-speaker intervals like \texttt{B1} system, and we propose using guided speaker embedding (\texttt{P1} in \autoref{tbl:result_veri}) instead.
For the proposed method, we input all the speech intervals with the corresponding speech activities to the extractor, i.e., silence intervals are excluded from the input.
Note that the speaker embedding extractor of the baseline was replaced with the one used in \texttt{B1} for fair comparisons because the speaker embedding extractor used in the original pyannote employs a different architecture (i.e., ResNet-34~\cite{he2016deep}) and training strategy (including speed perturbation and large-margin fine-tuning~\cite{thienpondt2021idlab}).
As in pyannote, we used agglomerative hierarchical clustering with centroid linkage to estimate the global speaker labels. 
The clustering threshold and the minimum cluster size were tuned for each dataset using the corresponding development portion.
We also validated the performance assuming that the oracle diarization result for each \SI{10}{\second} with \SI{1}{\second} shift is assumed to be obtained.
The evaluation metrics to measure the performance are diarization error rate (DER) and Jaccard error rate (JER) with no forgiveness of collar.
In short, DER is more influenced by dominant speakers, while JER treats each speaker equally in its calculation and is affected by the less frequently speaking speakers, which are difficult to predict correctly.

The results of speaker diarization with the baseline and guided speaker embedding are shown in \autoref{tbl:result_diar}.
The proposed guided speaker embedding improved the DERs of pyannote 3.1 on six out of eight datasets by simply replacing the basic speaker embedding extractor with the proposed model.
Especially, the performance on the datasets with a high overlap ratio ($>$\SI{10}{\percent}) was consistently improved by using guided speaker embedding.
This indicated that a highly overlapped dataset resulted in short single-speaker intervals, raising the necessity of extracting speaker information from overlapped intervals.
In terms of JER, the proposed method always performed better than the baseline method.
This can be explained by the proposed method improving the discriminativeness of speaker embeddings in overlapping intervals, thus avoiding merging less frequently speaking speakers into the cluster of dominant speakers.
We also provided the results of the original pyannote for reference in \autoref{tbl:result_diar}.
It was 0.24/0.77 points better than \texttt{B1} in DER and JER overall, respectively, since the embedding extractor in the original pyannote used different architecture and training strategies as explained, but the proposed method still performed better.
Incorporating more advanced training strategies into the proposed method is left for future work.

The same trend was observed when the oracle local diarization results were used.
The proposed method showed better DERs on five out of eight datasets, including the datasets of top-4 overlap ratios, and the gaps between the baseline and the proposed method were quite small on the three datasets in which the baseline showed better DERs.
Also, in terms of JER, the proposed method outperformed the baseline on all the datasets except for AISHELL-4.
These findings indicate that the investigation of speaker embeddings can be carried out independently from local diarization and that the proposed method is still effective with perfect local diarization.

\section{Conclusion}
This paper proposed guided speaker embedding, which uses speaker activities as clues to extract the target speaker's embedding from multi-speaker recordings.
In this method, speaker activities are concatenated to input acoustic features to suppress the information of interference speakers in the encoder, and are also used to mask the attention weights to focus only on the intervals in which the target speaker is active.
The effectiveness of the proposed method was validated under speaker verification and diarization.
Future work will include a joint investigation of the proposed method and multi-speaker ASR.

\clearpage
\IEEEtriggeratref{23}
\bibliographystyle{IEEEbib}
\bibliography{mybib}

\begin{thebibliography}{10}

\bibitem{lu2021streaming}
Liang Lu, Naoyuki Kanda, Jinyu Li, and Yifan Gong,
\newblock ``Streaming end-to-end multi-talker speech recognition,''
\newblock {\em IEEE Signal Processing Letters}, vol. 28, pp. 803--807, 2021.

\bibitem{lu2022endpoint}
Liang Lu, Jinyu Li, and Yifan Gong,
\newblock ``Endpoint detection for streaming end-to-end multi-talker {ASR},''
\newblock in {\em Proc. ICASSP}, 2022, pp. 7312--7316.

\bibitem{sklyar2022multi}
Ilya Sklyar, Anna Piunova, Xianrui Zheng, and Yulan Liu,
\newblock ``Multi-turn {RNN-T} for streaming recognition of multi-party speech,''
\newblock in {\em Proc. ICASSP}, 2022, pp. 8402--8406.

\bibitem{kanda2022streaming}
Naoyuki Kanda, Jian Wu, Yu~Wu, Xiong Xiao, Zhong Meng, Xiaofei Wang, Yashesh Gaur, Zhuo Chen, Jinyu Li, and Takuya Yoshioka,
\newblock ``Streaming multi-talker {ASR} with token-level serialized output training,''
\newblock in {\em Proc. Interspeech}, 2022, pp. 3774--3778.

\bibitem{moriya2025alignment}
Takafumi Moriya, Shota Horiguchi, Marc Delcroix, Ryo Masumura, Takanori Ashihara, Hiroshi Sato, Kohei Matsuura, and Masato Mimura,
\newblock ``Alignment-free training for transducer-based multi-talker {ASR},''
\newblock in {\em Proc. ICASSP}, 2025.

\bibitem{kinoshita2021integrating}
Keisuke Kinoshita, Marc Delcroix, and Naohiro Tawara,
\newblock ``Integrating end-to-end neural and clustering-based diarization: Getting the best of both worlds,''
\newblock in {\em Proc. ICASSP}, 2021, pp. 7198--7202.

\bibitem{kinoshita2021advances}
Keisuke Kinoshita, Marc Delcroix, and Naohiro Tawara,
\newblock ``Advances in integration of end-to-end neural and clustering-based diarization for real conversational speech,''
\newblock in {\em Proc. INTERSPEECH}, 2021, pp. 3565--3569.

\bibitem{bredin2023pyannote}
Herv{\'e} Bredin,
\newblock ``{pyannote.audio 2.1} speaker diarization pipeline: principle, benchmark, and recipe,''
\newblock in {\em Proc. Interspeech}, 2023, pp. 1983--1987.

\bibitem{plaquet2023powerset}
Alexis Plaquet and Herv{\'e} Bredin,
\newblock ``Powerset multi-class cross entropy loss for neural speaker diarization,''
\newblock in {\em Proc. Interspeech}, 2023, pp. 3222--3226.

\bibitem{lu2021streaming2}
Liang Lu, Naoyuki Kanda, Jinyu Li, and Yifan Gong,
\newblock ``Streaming multi-talker speech recognition with joint speaker identification,''
\newblock in {\em Proc. Interspeech}, 2021, pp. 1782--1786.

\bibitem{kanda2022streaming2}
Naoyuki Kanda, Jian Wu, Yu~Wu, Xiong Xiao, Zhong Meng, Xiaofei Wang, Yashesh Gaur, Zhuo Chen, Jinyu Li, and Takuya Yoshioka,
\newblock ``Streaming speaker-attributed {ASR} with token-level speaker embeddings,''
\newblock in {\em Proc. Interspeech}, 2022, pp. 521--525.

\bibitem{tawara2024ntt}
Naohiro Tawara, Marc Delcroix, Atsushi Ando, and Atsunori Ogawa,
\newblock ``{NTT} speaker diarization system for {CHiME-7}: Multi-domain, multi-microphone end-to-end and vector clustering diarization,''
\newblock in {\em Proc. ICASSP}, 2024, pp. 11281--11285.

\bibitem{nagrani2020voxceleb}
Arsha Nagrani, Joon~Son Chung, Weidi Xie, and Andrew Zisserman,
\newblock ``{VoxCeleb}: Large-scale speaker verification in the wild,''
\newblock {\em Computer Speech \& Language}, vol. 60, pp. 101027, 2020.

\bibitem{snyder2017deep}
David Snyder, Daniel Garcia-Romero, Daniel Povey, and Sanjeev Khudanpur,
\newblock ``Deep neural network embeddings for text-independent speaker verification,''
\newblock in {\em Proc. Interspeech}, 2017, pp. 999--1003.

\bibitem{snyder2018xvectors}
David Snyder, Daniel Garcia-Romero, Grregory Sell, Daniel Povey, and Sanjeev Khudanpur,
\newblock ``X-vectors: Robust {DNN} embeddings for speaker recognition,''
\newblock in {\em Proc. ICASSP}, 2018, pp. 5329--5333.

\bibitem{desplanques2020ecapatdnn}
Brecht Desplanques, Jenthe Thienpondt, and Kris Demuynck,
\newblock ``{ECAPA-TDNN}: Emphasized channel attention, propagation and aggregation in {TDNN} based speaker verification,''
\newblock in {\em Proc. Interspeech}, 2020, pp. 3830--3834.

\bibitem{zhou2021resnext}
Tianyan Zhou, Yong Zhao, and Jian Wu,
\newblock ``{ResNeXt} and {Res2Net} structures for speaker verification,''
\newblock in {\em Proc. SLT}, 2021, pp. 301--307.

\bibitem{han2020mirnet}
Hyewon Han, Soo-Whan Chung, and Hong-Goo Kang,
\newblock ``{MIRNet}: Learning multiple identities representations in overlapped speech,''
\newblock in {\em Proc. Interspeech}, 2020, pp. 4303--4307.

\bibitem{cord2023teacher}
Tobias Cord-Landwehr, Christoph Boeddeker, C{\u{a}}t{\u{a}}lin Zoril{\u{a}}, Rama Doddipatla, and Reinhold Haeb-Umbach,
\newblock ``A teacher-student approach for extracting informative speaker embeddings from speech mixtures,''
\newblock in {\em Proc. Interspeech}, 2023, pp. 4703--4707.

\bibitem{horiguchi2024recursive}
Shota Horiguchi, Atsushi Ando, Takafumi Moriya, Takanori Ashihara, Hiroshi Sato, Naohiro Tawara, and Marc Delcroix,
\newblock ``Recursive attentive pooling for extracting speaker embeddings from multi-speaker recordings,''
\newblock in {\em Proc. SLT}, 2024, pp. 1219--1226.

\bibitem{taherian2024multi}
Hassan Taherian and DeLiang Wang,
\newblock ``Multi-channel conversational speaker separation via neural diarization,''
\newblock {\em IEEE/ACM TASLP}, vol. 32, pp. 2467--2476, 2024.

\bibitem{kanda2019simultaneous}
Naoyuki Kanda, Shota Horiguchi, Yusuke Fujita, Yawen Xue, Kenji Nagamatsu, and Shinji Watanabe,
\newblock ``Simultaneous speech recognition and speaker diarization for monaural dialogue recordings with target-speaker acoustic models,''
\newblock in {\em Proc. ASRU}, 2019, pp. 31--38.

\bibitem{medennikov2020targetspeaker}
Ivan Medennikov, Maxim Korenevsky, Tatiana Prisyach, Yuri Khokhlov, Mariya Korenevskaya, Ivan Sorokin, Tatiana Timofeeva, Anton Mitrofanov, Andrei Andrusenko, Ivan Podluzhny, Aleksandr Laptev, and Aleksei Romanenko,
\newblock ``Target-speaker voice activity detection: a novel approach for multi-speaker diarization in a dinner party scenario,''
\newblock in {\em Proc. INTERSPEECH}, 2020, pp. 274--278.

\bibitem{boeddeker2018front}
Christoph Boeddeker, Jens Heitkaemper, Joerg Schmalenstoeer, Lukas Drude, Jahn Heymann, and Reinhold Haeb-Umbach,
\newblock ``Front-end processing for the {CHiME-5} dinner party scenario,''
\newblock in {\em Proc. CHiME-5}, 2018, pp. 35--40.

\bibitem{watanabe2020chime}
Shinji Watanabe, Michael Mandel, Jon Barker, Emmanuel Vincent, Ashish Arora, Xuankai Chang, Sanjeev Khudanpur, Vimal Manohar, Daniel Povey, Desh Raj, David Snyder, Aswin~Shanmugam Subramanian, Jan Trmal, Bar~Ben Yair, Christoph Boeddeker, Zhaoheng Ni, Yusuke Fujita, Shota Horiguchi, Naoyuki Kanda, Takuya Yoshioka, and Neville Ryant,
\newblock ``{CHiME-6 Challenge}: Tackling multispeaker speech recognition for unsegmented recordings,''
\newblock in {\em Proc. CHiME-6}, 2020.

\bibitem{he2023ansdmamse}
Mao-Kui He, Jun Du, Qing-Feng Liu, and Chin-Hui Lee,
\newblock ``{ANSD-MA-MSE}: Adaptive neural speaker diarization using memory-aware multi-speaker embedding,''
\newblock {\em IEEE/ACM TASLP}, vol. 31, pp. 1561--1573, 2023.

\bibitem{yang2024neural}
Gaobin Yang, Maokui He, Shutong Niu, Ruoyu Wang, Yanyan Yue, Shuangqing Qian, Shilong Wu, Jun Du, and Chin-Hui Lee,
\newblock ``Neural speaker diarization using memory-aware multi-speaker embedding with sequence-to-sequence architecture,''
\newblock in {\em Proc. ICASSP}, 2024, pp. 11626--11630.

\bibitem{delcroix2021speaker}
Marc Delcroix, Katerina Zmolikova, Tsubasa Ochiai, Keisuke Kinoshita, and Tomohiro Nakatani,
\newblock ``Speaker activity driven neural speech extraction,''
\newblock in {\em Proc. ICASSP}, 2021, pp. 6099--6103.

\bibitem{snyder2015musan}
David Snyder, Guoguo Chen, and Daniel Povey,
\newblock ``{MUSAN}: A music, speech, and noise corpus,'' arXiv:1510.08484, 2015.

\bibitem{ko2017study}
Tom Ko, Vijayaditya Peddinti, Daniel Povey, Michael~L Seltzer, and Sanjeev Khudanpur,
\newblock ``A study on data augmentation of reverberant speech for robust speech recognition,''
\newblock in {\em Proc. ICASSP}, 2017, pp. 5220--5224.

\bibitem{fu2021aishell}
Yihui Fu, Luyao Cheng, Shubo Lv, Yukai Jv, Yuxiang Kong, Zhuo Chen, Yanxin Hu, Lei Xie, Jian Wu, Hui Bu, Xin Xu, Jun Du, and Jingdong Chen,
\newblock ``{AISHELL-4}: An open source dataset for speech enhancement, separation, recognition and speaker diarization in conference scenario,''
\newblock in {\em Proc. Interspeech}, 2021, pp. 3665--3669.

\bibitem{yu2022m2met}
Fan Yu, Shiliang Zhang, Yihui Fu, Lei Xie, Siqi Zheng, Zhihao Du, Weilong Huang, Pengcheng Guo, Zhijie Yan, Bin Ma, Xin Xu, and Hui Bu,
\newblock ``{M2MeT}: The {ICASSP} 2022 multi-channel multi-party meeting transcription challenge,''
\newblock in {\em Proc. ICASSP}, 2022, pp. 6167--6171.

\bibitem{carletta2007unleashing}
Jean Carletta,
\newblock ``Unleashing the killer corpus: experiences in creating the multi-everything {AMI Meeting Corpus},''
\newblock {\em Language Resources and Evaluation}, vol. 41, no. 2, pp. 181--190, 2007.

\bibitem{liu2022msdwild}
Tao Liu, Shuai Fan, Xu~Xiang, Hongbo Song, Shaoxiong Lin, Jiaqi Sun, Tianyuan Han, Siyuan Chen, Binwei Yao, Sen Liu, Yifei Wu, Yanmin Qian, and Kai Yu,
\newblock ``{MSDWild}: Multi-modal speaker diarization dataset in the wild,''
\newblock in {\em Proc. Interspeech}, 2022, pp. 1476--1480.

\bibitem{ryant2021third}
Neville Ryant, Prachi Singh, Venkat Krishnamohan, Rajat Varma, Kenneth Church, Christopher Cieri, Jun Du, Sriram Ganapathy, and Mark Liberman,
\newblock ``The third {DIHARD} diarization challenge,''
\newblock in {\em Proc. INTERSPEECH}, 2021, pp. 3570--3574.

\bibitem{chung2020spot}
Joon~Son Chung, Jaesung Huh, Arsha Nagrani, Triantafyllos Afouras, and Andrew Zisserman,
\newblock ``Spot the conversation: Speaker diarisation in the wild,''
\newblock in {\em Proc. Interspeech}, 2020, pp. 299--303.

\bibitem{bredin2020pyannote}
Herv{\'e} Bredin, Ruiqing Yin, Juan~Manuel Coria, Gregory Gelly, Pavel Korshunov, Marvin Lavechin, Diego Fustes, Hadrien Titeux, Wassim Bouaziz, and Marie-Philippe Gill,
\newblock ``\texttt{pyannote.audio}: neural building blocks for speaker diarization,''
\newblock in {\em Proc. ICASSP}, 2020, pp. 7124--7128.

\bibitem{he2016deep}
Kaiming He, Xiangyu Zhang, Shaoqing Ren, and Jian Sun,
\newblock ``Deep residual learning for image recognition,''
\newblock in {\em Proc. CVPR}, 2016, pp. 770--778.

\bibitem{thienpondt2021idlab}
Jenthe Thienpondt, Brecht Desplanques, and Kris Demuynck,
\newblock ``The {IDLab} {VoxSRC}-20 submission: Large margin fine-tuning and quality-aware score calibration in {DNN} based speaker verification,''
\newblock in {\em Proc. ICASSP}, 2021, pp. 5814--5818.

\end{thebibliography}

\end{document}